\documentclass[a4paper,11pt]{article}
\usepackage{pos_mod}
\usepackage{colortbl}
\usepackage{multirow}

\newcommand{\bq}{\begin{eqnarray}}
\newcommand{\eq}{\end{eqnarray}}
\newcommand{\eps}{\varepsilon}

\newcommand{\loopnumber}{l}
\newcommand{\nedges}{n_{\mathrm{edges}}}

\newcommand{\NV}{n}

\newcommand{\ND}{N_D}

\newcommand{\Divisor}{P}
\newcommand{\divisor}{p}

\newcommand{\differentialform}{\Psi}

\newcommand{\Fcomb}{F_{\mathrm{comb}}}
\newcommand{\Fgeom}{F_{\mathrm{geom}}}
\newcommand{\Fgen}{F}

\newcommand{\Acomb}{\Omega_{\mathrm{comb}}}
\newcommand{\Ageom}{\Omega_{\mathrm{geom}}}
\newcommand{\Agen}{\Omega}

\newcommand{\Hcomb}{H_{\mathrm{comb}}}
\newcommand{\Hgeom}{H_{\mathrm{geom}}}
\newcommand{\Hgen}{H}

\newcommand{\hcomb}{h_{\mathrm{comb}}}
\newcommand{\hgeom}{h_{\mathrm{geom}}}

\newcommand{\absmu}{|\mu|}

\newcommand{\laportaorder}{(a,w,o,|\mu|,\dots)}

\title{Improving integration-by-parts and differential equations}
\ShortTitle{Improving IBPs and DEs}

\author[a]{Iris~Bree}
\author[b]{Federico~Gasparotto}
\author[a]{Antonela~Matija\v{s}i\'c}
\author[a]{Pouria~Mazloumi}
\author[a]{Dmytro~Melnichenko}
\author[c]{Sebastian~P\"ogel}
\author[a]{Toni~Teschke}
\author[d]{Xing~Wang}
\author*[a]{Stefan~Weinzierl}
\author[e]{Konglong~Wu}
\author[f]{Xiaofeng~Xu}

\affiliation[a]{PRISMA Cluster of Excellence, Universit{\"a}t Mainz, 55099 Mainz, Germany}

\affiliation[b]{Bethe Center for Theoretical Physics, Universität Bonn, 53115 Bonn, Germany}

\affiliation[c]{Department of Astrophysics, University of Zurich, Winterthurerstrasse 190, 8057 Zurich, Switzerland}

\affiliation[d]{School of Science and Engineering, The Chinese University of Hong Kong, Shenzhen, 518172 Guangdong, China}

\affiliation[e]{School of Physics and Technology, Wuhan University, No.299 Bayi Road, Wuhan 430072, China}

\affiliation[f]{Department of Physics, Xiamen University, Xiamen, 361005, China}

\emailAdd{weinzierl@uni-mainz.de}

\abstract{
In this talk, we discuss how ideas from geometry help to improve Feynman integral reduction and the construction of $\varepsilon$-factorised differential equations. 
In particular, we outline a systematic procedure to obtain an $\varepsilon$-factorised differential equation for any Feynman integral.
}

\FullConference{17th International Symposium on Radiative Corrections: Applications of Quantum Field Theory to Phenomenology (RADCOR2025)\\
5-10 October 2025\\
Puri, India\\}

\begin{document}
\maketitle


\section{Introduction}

Experiments in high-energy physics, like the ones at the LHC and future colliders, rely on precision
and require input from the theory side in the form of precision calculations.
Perturbative quantum field theory and, in consequence, Feynman integrals, are the tools for theoretical precision calculations.
Feynman integrals are not only relevant to high-energy physics, but also find applications in gravitational wave physics and low-energy precision experiments.
Current research focuses therefore on advancing the techniques for these calculations \cite{Chen:2020uyk,Chen:2022lzr,DHoker:2023khh,Marzucca:2023gto,delaCruz:2024xit,Baune:2024biq,Baune:2024ber,Jockers:2024uan,Gehrmann:2024tds,Pogel:2024sdi,Duhr:2024xsy,Gasparotto:2024bku,DHoker:2025szl,DHoker:2025dhv,Duhr:2025ppd,Duhr:2025tdf,Chaubey:2025adn,Bargiela:2025vwl,Carrolo:2026qpu}.
Perturbative calculations are carried out with the help of computer algebra and
rely on a few basic algorithms for Feynman integral reduction and the computation of master integrals.
These algorithms require significant computing resources in the form of memory and CPU time and are therefore often the bottleneck.
Improving the efficiency of these basic algorithms will have a direct impact on the higher-order calculations which can be done.

One of the basic algorithms is Feynman integral reduction.
It is based on integration-by-parts identities \cite{Tkachov:1981wb,Chetyrkin:1981qh} and the Laporta algorithm \cite{Laporta:2000dsw}.
There are several public computer program implementations \cite{vonManteuffel:2012np,Smirnov:2014hma,Maierhoefer:2017hyi,Wu:2023upw,Guan:2024byi}.

A second basic algorithm is the computation of master integrals 
through the method of differential equations~\cite{Kotikov:1990kg,Kotikov:1991pm,Remiddi:1997ny,Gehrmann:1999as,Henn:2013pwa}.
This method can be used analytically or numerically.
One utilises integration-by-parts identities to derive 
a set of (non-$\eps$-factorised) differential equations.
This step is algorithmic and involves only linear algebra.
The only limitation is the availability of computing resources. 
For an analytic calculation one usually performs two additional steps: 
In the second step, one transforms the system of differential equations to an $\eps$-factorised form \cite{Henn:2013pwa}.
In the last step, one solves the $\eps$-factorised differential equations order by order in $\eps$ in terms of iterated integrals \cite{Chen}.
The third step is also straightforward, and there are no conceptual issues, provided appropriate boundary values are given.
Since the boundary values depend on one fewer kinematic variable, they are simpler to calculate. In fact,
they can be recursively reduced to single-mass vacuum integrals \cite{Liu:2022chg,Liu:2017jxz,Liu:2022mfb}.
Analytic calculations based on the above-mentioned $\eps$-factorised form have been successfully employed 
in several state-of-the-art calculations, including calculations with non-trivial geometries \cite{Henn:2025xrc,Adams:2018yfj,Honemann:2018mrb,Bogner:2019lfa,Muller:2022gec,Giroux:2022wav,Dlapa:2022wdu,Gorges:2023zgv,Delto:2023kqv,Jiang:2023jmk, Ahmed:2024tsg,Giroux:2024yxu,Duhr:2024bzt,Schwanemann:2024kbg,Marzucca:2025eak,Becchetti:2025oyb,Becchetti:2025rrz, Ahmed:2025osb, Chen:2025hzq,Coro:2025vgn,Pogel:2022vat,Pogel:2022yat,Pogel:2022ken,Duhr:2022dxb,Forner:2024ojj,Frellesvig:2024rea,Duhr:2025lbz,Maggio:2025jel,Duhr:2025kkq,Pogel:2025bca,Duhr:2024uid}.

Despite this success, there are two bottlenecks:
The first bottleneck is the availability of computing resources for the required integration-by-parts reduction.
The second bottleneck is conceptual:
Can one always find a transformation to an $\eps$-factorised differential equation?
In this talk we discuss how to improve the two basic algorithms.

\section{Integration-by-parts}

\begin{figure}
\begin{center}
\includegraphics[scale=1.0]{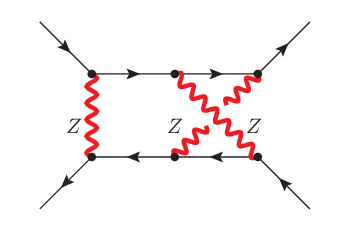}
\end{center}
\caption{
A non-planar double box integral.
}
\label{fig:moeller}
\end{figure}

Feynman integral reduction with standard implementations of the Laporta algorithm may introduce spurious polynomials
in the denominator. These spurious polynomials lead to an expression growth, consume memory and slow the reduction programs down.
In order to see the problem, consider the following linear system of equations:
\bq
\begin{array}{rrrrcl}
 14732 \; x_1 & - 2514 \; x_2 & - 5 \; x_3 & - 7 \; x_4 & = & 0, \\
 9872 \; x_1 & - 17294 \; x_2 & + 3 \; x_3 & - 11 \; x_4 & = & 0, \\
 5068 \; x_1 & - 49336 \; x_2 & + 18 \; x_3 & - 22 \; x_4 & = & 0. \\
\end{array}
\nonumber 
\eq
This system has rank $2$.
If we solve for $x_1$ and $x_2$ and take $x_3$ and $x_4$ as free variables we obtain
\bq
\label{eq_1}
 x_1 \;= \; \frac{1237}{3025750} x_3 + \frac{1229}{3025750} x_4,
 & &
 x_2 \; = \; \frac{1231}{3025750} x_3 - \frac{1223}{3025750} x_4.
\eq
We note a large denominator.
However, if we solve for $x_3$ and $x_4$ and take $x_1$ and $x_2$ as free variables we find
\bq
\label{eq_2}
 x_3 \; = \; 1223 x_1 + 1229 x_2,
 & & 
 x_4 \; = \; 1231 x_1 - 1237 x_2.
\eq
This set of equations is simpler and does not involve a large denominator.
Our interest is to avoid large denominators.
In the context of Feynman integrals, the variables $x_1-x_4$ correspond to Feynman integrals and the free variables to master integrals.
Large denominators are lengthy polynomials in the kinematic variables and the dimensional regularisation parameter.
Such spurious polynomials often occur in connection with sectors with a high number of master integrals and non-trivial geometries.
There are some heuristic methods which try to avoid the occurrence of spurious polynomials \cite{Smirnov:2020quc,Usovitsch:2020jrk}.
An example which may lead to spurious polynomials is shown in fig.~\ref{fig:moeller}.
This Feynman integral contributes to two-loop electro-weak corrections 
for Bhabha scattering, M{\o}ller scattering and Drell-Yan.
It depends on the Mandelstam variables $s$ and $t$ and an internal mass $m_Z$.
The family of this Feynman integral has $67$ master integrals.
It is not an innocent example.
The top sector with $5$ master integrals corresponds to a curve of genus two.
In addition, there is a sub-sector with $8$ master integrals corresponding to a K3-surface
and five sub-sectors corresponding to five different elliptic curves.

The Laporta algorithm uses Gau{\ss} elimination and an order relation.
The order relation determines the free variables (master integrals).
In eq.~(\ref{eq_1}) the variables $x_1,x_2$ are considered more complicated than $x_3,x_4$, in eq.~(\ref{eq_2}) it is the other way round.
Up to now, standard Feynman integral reduction programs order Feynman integrals by sectors, but within a sector only an ad-hoc order is used.
Typical order relations are the lexicographical order of tuples
\bq
\mbox{ISP-basis}: & & \left( N_{\mathrm{prop}}, N_{\mathrm{id}}, r_{\mathrm{dot}}, s_{\mathrm{ISP}}, \dots \right),
 \nonumber \\
\mbox{dot-basis}: & & \left( N_{\mathrm{prop}}, N_{\mathrm{id}}, s_{\mathrm{ISP}}, r_{\mathrm{dot}}, \dots \right),
 \nonumber
\eq
where $N_{\mathrm{prop}}$ denotes the number of propagators, $N_{\mathrm{id}}$ the sector id, $r_{\mathrm{dot}}$ the number of dots, $s_{\mathrm{ISP}}$
the number of irreducible scalar products in the numerator and the dots stand for further criteria needed to distinguish inequivalent integrals.
In these order relations the ordering within one sector is determined by $(\dots,r_{\mathrm{dot}}, s_{\mathrm{ISP}}, \dots )$ or $(\dots,s_{\mathrm{ISP}}, r_{\mathrm{dot}}, \dots )$, respectively. 
While the number of master integrals is independent of the chosen order relation, the size of the reduction tables may depend on the choice.

This raises the question, if we can improve the efficiency of the Laporta algorithm by choosing a better order relation.
A second related question asks, if we can decompose the vector space of master integrals on the maximal cut in a meaningful way 
into smaller spaces.
Both questions have a positive answer \cite{e-collaboration:2025frv,Bree:2025tug}.
In perturbative calculations the unknown variables are Feynman integrals and we have additional information on them.
In particular, we can associate to each Feynman integral a geometry and we may use an order relation motivated by this geometry.

\begin{figure}
\begin{center}
\includegraphics[scale=1.0]{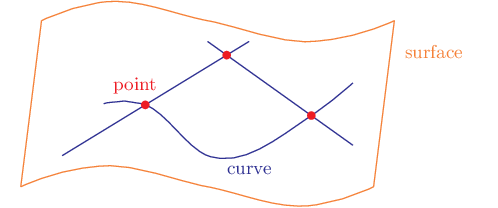}
\end{center}
\caption{
A mixed geometry: Inside a surface of dimension two there can be curves of dimension one and points of dimension zero.
}
\label{fig:mixed_geometry}
\end{figure}

To a first approximation the geometry of a Feynman integral is determined by the maximal cut of the Feynman integral.
The maximal cut of a Feynman integral is most easily analysed in the loop-by-loop Baikov representation \cite{Frellesvig:2017aai}.
On the maximal cut we define the 
Baikov polynomials $p_i(z)$ by
\bq
 \int\limits_{{\mathcal C}_{\mathrm{maxcut}}} \prod\limits_{r=1}^{\loopnumber} \frac{d^Dk_r}{i \pi^{\frac{D}{2}}} 
 \frac{1}{\prod\limits_{j=1}^{\nedges} \sigma_j}
 & \sim & 
 \int d^{\NV}z \;
 \prod\limits_{i \in I_{\mathrm{all}}} \left[ \divisor_i\left(z\right) \right]^{\alpha_i}.
\eq
The exponents $\alpha_i$ are always of the form
\bq
 \alpha_i \; = \; 
 \frac{1}{2} \left( a_i + b_i \eps \right),
 & \mbox{with} &
 a_i,b_i \; \in \; {\mathbb Z}.
\eq
We require the representation to be a good Baikov representation \cite{Bree:2025tug}, i.e. all singularities are regularised by the twist.
We define $I_{\mathrm{odd}}$ as the set of indices for which $a_i$ is odd
and $I_{\mathrm{even}}$ as the set of indices for which $a_i$ is even.
Instead of working in the affine chart $z=(z_1,\dots,z_{\NV})$, we go to projective space ${\mathbb C}{\mathbb P}^{\NV}$
with homogeneous coordinates $[z_0:z_1:\dots:z_{\NV}]$.
We let $P_i$ be the homogenisation of $p_i$.
In order to maintain homogeneity we introduce $P_0=z_0$ with exponent $\alpha_0 = \frac{1}{2}(a_0+b_0\eps)$.
We can unify the notation by including the index $0$ in $I_{\mathrm{even}}$ or $I_{\mathrm{odd}}$, 
depending on $a_0$ being even or odd, respectively.
We denote the resulting index sets by $I_{\mathrm{even}}^0$, $I_{\mathrm{odd}}^0$ and $I_{\mathrm{all}}^0$.
The integrands of Feynman integrals can be viewed as twisted cohomology classes \cite{Mastrolia:2018uzb,Frellesvig:2019uqt}.
Within twisted cohomology, we may always move integer powers of the Baikov polynomials between the twist function 
and the rational differential $\NV$-form.
We can therefore define a ``minimal'' twist function, by requiring $a_i \in \{-1,0\}$ for all $i$:
\bq
 U\left(z_0,z_1,\dots,z_{\NV}\right)
 & = &
 \prod\limits_{i \in I_{\mathrm{odd}}^0} \Divisor_i^{-\frac{1}{2}+\frac{1}{2} b_i \eps}
 \prod\limits_{j \in I_{\mathrm{even}}^0} \Divisor_j^{\frac{1}{2} b_j \eps}.
\eq
We therefore study integrands of the form
\bq
\label{objects}
 \differentialform_{\mu_0 \dots \mu_{\ND}}\left[Q\right]
 = 
 \mbox{prefactor}
 \cdot 
 \mbox{twist}
 \; \cdot
 \underbrace{\frac{Q}{\prod\limits_{i \in I_{\mathrm{all}}^0} \Divisor_i^{\mu_i}}}_{\mathrm{rational} \; \mathrm{function}}
 \cdot \;
 \mbox{standard $\NV$-form}.
\eq
We denote the vector space spanned by the differential forms of eq.~(\ref{objects}) by $\Agen^{\NV}$.
The even and the odd polynomials will play different roles.
If an even polynomial is present in the denominator, we may take a residue and reduce to a simpler problem of dimension $(\NV-1)$.
The odd polynomials define a geometry
\bq
\label{def_hypersurface_square_free}
 y^2 - \prod\limits_{i \in I^0_{\mathrm{odd}}} \Divisor_i\left(z\right) & = & 0
\eq
in a suitable weighted projective space.
On the maximal cut we have in general a mixed geometry: Inside the geometry of dimension $\NV$ defined by eq.~(\ref{def_hypersurface_square_free})
there can be sub-geometries of dimension $(\NV-1)$, which in turn might have sub-sub-geometries of dimension $(\NV-2)$, etc.
This is shown in fig.~\ref{fig:mixed_geometry}.
This is similar to the amplituhedron \cite{Arkani-Hamed:2013jha,Arkani-Hamed:2017tmz}, the essential difference is that for Feynman integrals we may have non-trivial differential forms 
(i.e. differential forms which are not dlog-forms) of order $k>0$.
These correspond to elliptic curves and more complicated geometries.

To each differential form $\differentialform_{\mu_0 \dots \mu_{\ND}}[Q]$ as in eq.~(\ref{objects}) 
we associate three integer numbers $(r,o,\absmu)$.
The first integer $r$ denotes the largest number such that the $r$-fold residue of $\differentialform^0_{\mu_0 \dots \mu_{\ND}}[Q]$ is non-zero, 
where $\differentialform^0_{\mu_0 \dots \mu_{\ND}}[Q]$ is defined by setting $\eps=0$ in the twist function.
The integer $o$ denotes the pole order of $\differentialform^0_{\mu_0 \dots \mu_{\ND}}[Q]$.
The last integer is given by the sum of the indices in eq.~(\ref{objects})
\bq
 \left| \mu \right|
 & =
 \mu_0 + \dots + \mu_{\ND}.
\eq
It is convenient to set $w=\NV+r$. The variable $w$ denotes the Hodge weight.
If an integrand admits $r$ consecutive non-zero residues, it may be localised on a $(\NV-r)$-dimensional
sub-geometry. We always have $r \le \NV$, as the procedure of taking residues stops at points.

There are linear relations among the integrands defined in eq.~(\ref{objects}).
These are integration-by-parts identities, distribution identities and cancellation identities.
The last two originate from the way we represent the differential forms in eq.~(\ref{objects}):
The input data is a list of indices $(\mu_0 \dots \mu_{\ND})$ and numerator polynomial $Q$.
With a meticulous definition of the prefactors, the linear relations take the form
\begin{enumerate}
\item
Integration-by-parts identities:
\bq
 0 = 
 \frac{1}{\eps}
 \differentialform_{\mu_0 \dots \mu_i \dots \mu_{\ND}}\left[\partial_{z_j} Q_+\right]
 +
 \sum\limits_{i \in I_{\mathrm{all}}^0} 
 \differentialform_{\mu_0 \dots (\mu_i+1) \dots \mu_{\ND}}\left[Q_+ \cdot \left( \partial_{z_j} P_i \right) \right]
\eq

\item Distribution identities:
\bq
 \differentialform_{\mu_0 \dots \mu_{\ND}}\left[Q_1+Q_2\right]
 = 
 \differentialform_{\mu_0 \dots \mu_{\ND}}\left[Q_1\right]
 +
 \differentialform_{\mu_0 \dots \mu_{\ND}}\left[Q_2\right]
\eq

\item Cancellation identities:
\bq
\label{cancellation_id}
 \differentialform_{\mu_0 \dots (\mu_j+1) \dots \mu_{\ND}}\left[P_j \cdot Q\right]
 =
 \frac{1}{\eps}
 \left( \frac{a_j}{2} - \mu_j + {\color{red}{\frac{b_j}{2} \eps}} \right)
 \differentialform_{\mu_0 \dots \mu_j \dots \mu_{\ND}}\left[Q\right]
\eq
\end{enumerate}
The twisted cohomology $\Hgen^{\NV}$ is given as the vector space $\Agen^{\NV}$ modulo these linear relations.
A few comments are in order:
For the integration-by-parts identities and the distribution identities the powers of $\eps$ may be inferred from $\mu_0+\dots+\mu_{\ND}$.
The subsystem formed by these relations may be solved by setting $\eps=1$.
The coefficients in the solution will be monomials in $\eps$, the exponent can be restored from $\mu_0+\dots+\mu_{\ND}$.
This is only spoiled by the cancellation identities, we marked the offending term in eq.~(\ref{cancellation_id}) in red.
If we can ensure that in all cancellation identities the pivot element comes from the left-hand side of the cancellation identity, then
the coefficients in the solution of the linear system are always Laurent polynomials in $\eps$.

In a similar way we may define the twisted cohomology groups on the localisations, by restricting to the sub-geometries.
This can be done without introducing algebraic extensions (roots), the details are described in ref.~\cite{Bree:2025tug}.

We are now in a position to define an order relation on the space $\Agen^{\NV}$.
We do this by giving preference to differential forms, which correspond to master integrands on sub-geometries.
Thus the highest priority will be given to master integrands, which correspond to master integrands on points.
We assign an integer $a=-w=-2\NV$ to those.
The second highest priority will be given to master integrands, which correspond to master integrands on curves.
We assign an integer $a=-w=-2\NV+1$ to those.
This process is repeated until we reach the space $\Agen^{\NV}$ of dimension $\NV$.
Note that a non-zero value of $a$ is only assigned to differential forms, which correspond to master integrands on sub-geometries.
Differential forms, which can be localised on sub-geometries, but are not chosen as master integrands will have a value $a=0$.
In general, a geometry of dimension $\NV$ might have more than one sub-geometry of dimension $(\NV-1)$.
The sub-geometries of dimension $(\NV-1)$ might intersect on sub-sub-geometries of dimension $(\NV-2)$.
These intersections can provide relations among the master integrands of the sub-geometries of dimension $(\NV-1)$.
This is taken into account by a merging procedure, the details are described in ref.~\cite{Bree:2025tug}.

With these considerations we define the order relation for the Laporta algorithm on the space $\Agen^{\NV}$ as
\bq
 \laportaorder,
\eq 
where 
the dots stand for further criteria needed to distinguish inequivalent integrands.
The relation $a_1 < a_2$ implies $\differentialform_1 < \differentialform_2$, with ties broken by $w$, etc. 
The first variable $a$ gives preference to differential forms, which correspond to master integrands on sub-geometries.
Once we determined all master integrands from the sub-geometries, we fill up the remaining masters from the current geometry.
This means, we now would like to avoid differential forms, which have residues and can be localised.
The second variable $w$ vetoes those.
Within the current geometry we order the differential forms by their pole order.
In the Calabi-Yau case, this gives the highest precedence to the holomorphic differential form (of pole order zero).
Finally, ties within a certain pole order are broken by the fourth variable $\absmu$.

\begin{figure}
\begin{center}
\includegraphics[scale=1.0]{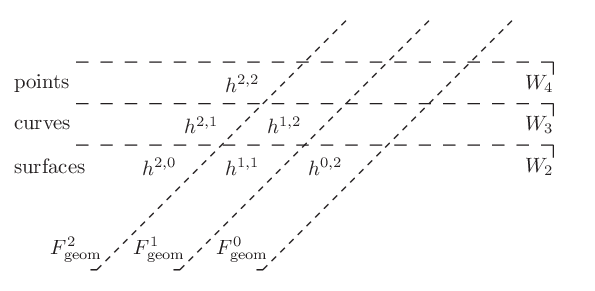}
\end{center}
\caption{
The decomposition of $\Hgeom^2$ into spaces $\Hgeom^{p,q}$ with the help of the filtrations $W_\bullet$ and $\Fgeom^\bullet$.
}
\label{fig:decomposition}
\end{figure}

\begin{figure}
\begin{center}
\begin{tabular}{lcl}
\begin{minipage}{2cm}
{
\vspace*{-30mm}
\flushright{3 masters}
}
\end{minipage}
&
\includegraphics[scale=0.6]{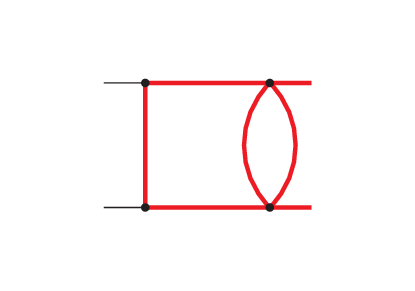}
&
\includegraphics[scale=0.6]{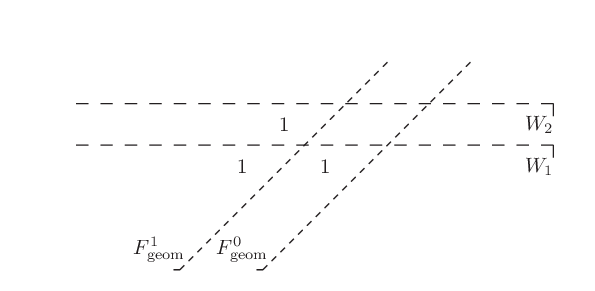}
\\
\begin{minipage}{2cm}
{
\vspace*{-30mm}
\flushright{5 masters}
}
\end{minipage}
&
\includegraphics[scale=0.6]{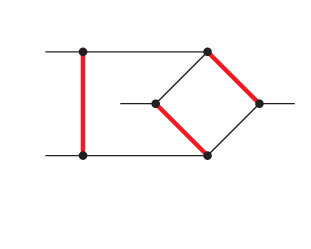}
&
\includegraphics[scale=0.6]{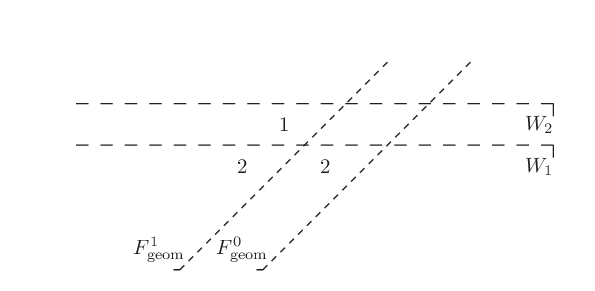}
\\
\begin{minipage}{2cm}
{
\vspace*{-38mm}
\flushright{11 masters}
}
\end{minipage}
&
\includegraphics[scale=0.6]{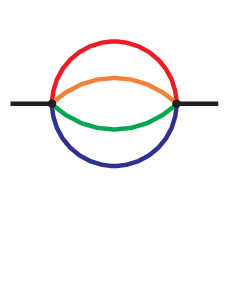}
&
\includegraphics[scale=0.6]{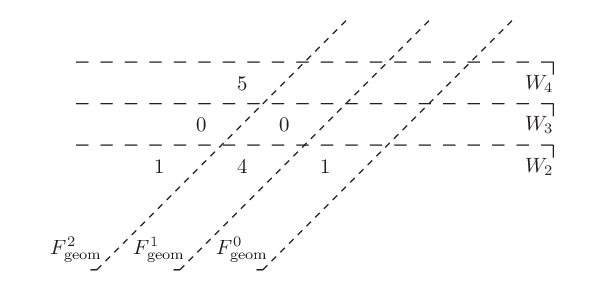}
\\
\end{tabular}
\end{center}
\caption{
Examples of decompositions of $V^{\NV}$ into spaces $V^{p,q}$ with the help of the filtrations $W_\bullet$ and $\Fgeom^\bullet$.
Coloured lines correspond to massive propagators.
}
\label{fig:examples_decomposition}
\end{figure}

We can be a little more formal and mathematical. On the vector space $\Agen^{\NV}$ we may introduce filtrations.
The general idea is that we filter a vector space by smaller and smaller subspaces.
\bq
 \mbox{filtration} & = & \mbox{something} + \mbox{simpler terms}.
 \nonumber 
\eq
Examples are a filtration defined by the number of consecutive non-zero residues 
\bq
 \mbox{filtration} & = & \mbox{term with non-zero residue} + \mbox{terms with zero residue},
 \nonumber
\eq
or a filtration defined by the pole order
\bq
 \mbox{filtration} & = & \mbox{term with highest pole order} + \mbox{terms with lower pole order}.
 \nonumber
\eq
Concretely, we define three filtrations.
The weight filtration $W_\bullet$ is defined by 
\begin{alignat}{2}
 \differentialform_{\mu_0 \dots \mu_{\ND}}[Q] & \in W_w \Agen^{\NV} & \quad \mbox{if} & \quad \NV + r \le w.
\end{alignat}
The weight filtration is the standard weight filtration from Hodge theory.
The filtration $\Fgeom^\bullet$ is defined by
\begin{alignat}{2}
 \differentialform_{\mu_0 \dots \mu_{\ND}}[Q] & \in \Fgeom^{p} \Agen^{\NV} & \quad \mbox{if} & \quad \NV+r-o \ge p.
\end{alignat}
At fixed weight $w$, the filtration $\Fgeom^\bullet$ is a filtration by the pole order $o$.
The third filtration $\Fcomb^\bullet$ is defined by 
\begin{alignat}{2}
 \differentialform_{\mu_0 \dots \mu_{\ND}}[Q] & \in \Fcomb^{p'} \Agen^{\NV} & \quad \mbox{if} & \quad \NV-\absmu \ge p'.
\end{alignat}
The combinatorial filtration $\Fcomb^\bullet$ is a filtration by the quantity $\absmu$.
The general idea is that we always work modulo simpler terms, i.e. modulo terms with fewer residues, lower pole order
or a smaller sum of indices $\absmu$. 
The following definitions formalise the concept of working modulo simpler terms: We set 
\bq
 \Ageom^{p,q} \; = \; \mathrm{Gr}^{p}_{\Fgeom} \mathrm{Gr}^W_{p+q} \Agen^{\NV}
 & \mbox{and} &
 \Acomb^{p',q'} \; = \; \mathrm{Gr}^{p'}_{\Fcomb} \mathrm{Gr}^W_{p'+q'} \Agen^{\NV},
\eq
where the graded parts are defined by 
\bq
 \mathrm{Gr}^W_{w} X \; = \; W_{w} X / W_{w-1} X
 & \mbox{and} &
 \mathrm{Gr}^{p}_{F} X \; = \; F^p X / F^{p+1} X.
\eq
For example, in $\Ageom^{p,q}$ we may ignore terms of weight $w'<p+q$ or terms with $p'>p$.
We carry over the filtrations to $\Hgen^{\NV}$ as follows:
We let $\Hgeom^{p,q}$ be generated by all master integrands $\differentialform \in \Ageom^{p,q}$
and $\Hcomb^{p',q'}$ be generated by all master integrands $\differentialform \in \Acomb^{p',q'}$.
We set 
\bq
 \hgeom^{p,q} \; = \; \dim \Hgeom^{p,q}
 & \mbox{and} &
 \hcomb^{p',q'} \; = \; \dim \Hcomb^{p',q'}.
\eq
We display this information in the form of a Hodge diagram, 
an example is shown for $\NV=2$ in fig.~\ref{fig:decomposition}.
Up to now we considered integrands, our ultimate interest is, however, the integrals.
We now discuss the relation between the two vector spaces.
We denote the vector space of Feynman integrals 
on the maximal cut modulo integration-by-parts identities by $V^{\NV}$, this is a finite-dimensional vector space.
There is an injective map
\bq
 \iota & : & V^{\NV} \rightarrowtail \Hgen^{\NV},
\eq
obtained from expressing the maximal cut of the Feynman integral in the Baikov representation.
It is clear that this map is injective: If the integrand of the Baikov representation is zero in $H^{\NV}$, 
then the maximal cut integral is zero in $V^{\NV}$.
However in general, the map will not be surjective.
There are two reasons for this:
First of all, integration can lead to symmetries among Feynman integrals (elements in $V^n$), which are not symmetries of the integrands (elements in $\Hgen^n$).
Secondly, a polynomial $\divisor_j(z)$ with $j \in I_{\mathrm{even}}$ can simply be a Baikov variable 
corresponding to an uncut inverse propagator.
In this case, $\Hgen^{\NV}$ will also contain the integrands of the sector 
where the exponent of this inverse propagator is positive.
If this sector has additional master integrals, they will also appear in $\Hgen^{\NV}$.
We call such a sector a super-sector.
Taking these subtleties into account, we can work in the space $\Hgen^{\NV}$ and convert back to $V^{\NV}$ in the end.
From the decomposition of $\Hgen^{\NV}$ based on the filtrations, we obtain a decomposition of $V^{\NV}$.
A few non-trivial examples are shown in fig.~\ref{fig:examples_decomposition}.
The first example has three master integrals on the maximal cut and appears in the two-loop corrections to
top-pair production \cite{Adams:2018bsn,Adams:2018kez}.
The loop-by-loop Baikov representation is one dimensional, and we encounter an elliptic curve in this example.
The integrand of the master integral in $V^{1,0}$ corresponds to the holomorphic one-form of the elliptic curve.
This is an Abelian differential of the first kind.
The integrand of the master integral in $V^{0,1}$ corresponds to an Abelian differential of the second kind
(with poles, but no residues).
Finally, the integrand of the master integral in $V^{1,1}$ corresponds 
to an Abelian differential of the third kind.
The story is essentially the same for the second example, except that we encounter in this example
a curve of genus two.
Hence, the two master integrals in $V^{1,0}$ correspond to the two holomorphic one-forms of the genus-2 curve
and the two master integrals in $V^{0,1}$ correspond to the two Abelian differentials of the second kind.
The third example shows a Feynman integral, where we encounter a K3 surface.

We may extend the order relation to the full system, including sub-sectors.
We use
\bq
\label{new_order_relation}
 \left( \hat{N}_{\mathrm{prop}}, \hat{N}_{\mathrm{id}}, a, w, o, \absmu, \dots \right),
\eq
where
\bq
 \left( \hat{N}_{\mathrm{prop}}, \hat{N}_{\mathrm{id}} \right)
 & = &
 \left\{ \begin{array}{ll}
  \left( N_{\mathrm{prop}}', N_{\mathrm{id}}' \right) & \mbox{if $N_{\mathrm{id}}'$ is the smallest sector such that $N_{\mathrm{id}}$ is a super-sector.} \\
  \left( N_{\mathrm{prop}}, N_{\mathrm{id}} \right) & \mbox{otherwise}. \\
 \end{array}
 \right.
 \nonumber 
\eq
This order relation performs significantly better than the standard order relation.
For the system of the non-planar double box integral shown in fig.~\ref{fig:moeller}
we find that the size (in bytes) of the differential equation for the master integrals is reduced by a factor of $25$.
The new order relation can be combined with existing optimisation techniques, like the use of finite fields \cite{Peraro:2016wsq,Peraro:2019svx}
or selecting a smaller set of relevant equations \cite{Guan:2024byi}.


\section{Differential equations}

The order relation in eq.~(\ref{new_order_relation}) has a second advantage:
We observe that 
the basis $J$ obtained from the Laporta algorithm with this order relation has the property that on the maximal cut
we have
\bq
\label{F_compatible}
 d J
 \; = \;
 A\left(\eps,x\right) J,
 & &
 A_{ij}\left(\eps,x\right)
 \; = \;
 \sum\limits_{k=-(\absmu_i-\absmu_j)}^1
 \eps^k A^{(k)}_{ij}\left(x\right).
\eq
We call such a differential equation an $\Fgen^\bullet$-compatible differential equation for the filtration $\Fcomb^\bullet$.
We illustrate this with an example, where we have three non-trivial parts in the filtration
\bq
 \emptyset = F^3 V \subseteq F^2 V \subseteq F^1 V \subseteq F^0 V = V.
\eq
In this case we can write 
\bq
 A & = & B^{(1)} + B^{(0)} + B^{(-1)} + B^{(-2)},
\eq
where
\begin{align}
\label{example_B}
 B^{(1)}
 &= 
 \left( \begin{array}{rrr}
 \eps A^{(1)}_{11} & \eps A^{(1)}_{12} & 0 \\ 
 \eps A^{(1)}_{21} & \eps A^{(1)}_{22} & \eps A^{(1)}_{23}  \\ 
 \eps A^{(1)}_{31} & \eps A^{(1)}_{32} & \eps A^{(1)}_{33}  \\ 
 \end{array} \right),
 &
 B^{(0)}
 & = 
 \left( \begin{array}{ccc}
 0 & 0 & 0 \\ 
 0 & 0 & 0  \\ 
 \cellcolor{green} A^{(0)}_{31} & 0 & 0   \\ 
 \end{array} \right),
 \nonumber \\
 B^{(-1)}
 & =
 \left( \begin{array}{ccc}
 0 & 0 & 0 \\ 
 \cellcolor{magenta} A^{(0)}_{21} & 0 & 0  \\ 
 \cellcolor{magenta} \frac{1}{\eps} A^{(-1)}_{31} & \cellcolor{cyan} A^{(0)}_{32} & 0   \\ 
 \end{array} \right),
 &
 B^{(-2)}
 & = 
 \left( \begin{array}{ccc}
 \cellcolor{yellow} A^{(0)}_{11} & 0 & 0 \\ 
 \cellcolor{yellow} \frac{1}{\eps} A^{(-1)}_{21} & \cellcolor{orange} A^{(0)}_{22} & 0  \\ 
 \cellcolor{yellow} \frac{1}{\eps^2} A^{(-2)}_{31} & \cellcolor{orange} \frac{1}{\eps} A^{(-1)}_{32} & \cellcolor{red} A^{(0)}_{33}   \\ 
 \end{array} \right).
\end{align}
In refs.~\cite{e-collaboration:2025frv,Bree:2025tug} we showed that from
a form as in eq.~(\ref{F_compatible}) we may always construct algorithmically a matrix $R_2$,
which leads to a basis $K=R_2^{-1}J$, such that 
the differential equation for $K$ is in $\eps$-factorised form.
In the example of eq.~(\ref{example_B}) we first remove the unwanted terms of $B^{(-2)}$, then the unwanted terms of $B^{(-1)}$ and finally the unwanted term of $B^{(0)}$.
Within a $B^{(k)}$ we proceed column-wise, starting with column one.
In general, the matrix $R_2$ will contain transcendental functions. 
These additional functions obey differential equations, which are similar, but simpler than the ones for the Feynman integrals we want to compute.

The procedure extends beyond the maximal cut. For sub-sectors we do not need to assume that the $\eps$-dependence of the terms 
in the non-diagonal blocks is given by a Laurent polynomial in $\eps$.
The algorithm can handle a rational dependence in $\eps$: 
One first performs a partial fraction decomposition in $\eps$ and then
removes any term which is not proportional to $\eps^1$.


\section{Conclusions}

In this talk we discussed improved algorithms for integration-by-parts reduction and the construction of $\eps$-factorised differential equations.
We propose to use in the Laporta algorithm a new order relation, which reflects the geometric properties of the Feynman integrals.

We observe that this order relation significantly improves the efficiency of integration-by-parts reduction.
We further observe that this order relation leads to a basis of master integrals, whose differential equation on the maximal cut is
compatible with a filtration.
We showed, that from a filtration-compatible differential equation we may always construct an $\eps$-factorised differential equation.
These improvements will be beneficial both to analytical approaches for the computation of Feynman integrals \cite{Chicherin:2020oor,Chicherin:2021dyp,Becchetti:2025rrz,Coro:2025vgn} 
as well as to (semi-) numerical approaches \cite{Hidding:2020ytt,Liu:2022chg,Liu:2017jxz,Liu:2022mfb,Armadillo:2022ugh,Prisco:2025wqs,PetitRosas:2025xhm}.

The methods discussed in this talk are notably independent of prior knowledge of the underlying geometry of the Feynman integral,
e.g. we don't need to know whether a particular Feynman integral corresponds to an elliptic curve, a K3 surface etc..
This is achieved through a more abstract language, which we borrowed from Hodge theory.

\subsection*{Acknowledgements}

This work has been supported by the Research Unit ``Modern Foundations of Scattering Amplitudes'' (FOR 5582)
funded by the German Research Foundation (DFG).
X.W. is supported by the University Development Fund of The Chinese University of Hong Kong, Shenzhen, under the Grant No. UDF01003912.
X.W. is also supported in part by the National Natural Science Foundation of China with Grant No. 12535006.
The work of F.G. is supported by the European Union (ERC Consolidator Grant LoCoMotive 101043686). Views and opinions
expressed are however those of the authors only and do not necessarily reflect those of the European Union or the European
Research Council. Neither the European Union nor the granting authority can be held responsible for them.

{\footnotesize
\bibliography{/home/stefanw/notes/biblio}
\bibliographystyle{h-physrev5}
}

\end{document}